# Proximity screening greatly enhances electronic quality of graphene

Daniil Domaretskiy[1], Zefei Wu[2], Van Huy Nguyen[2], Ned Hayward[1], Ian Babich[3,4], Xiao Li[2], Ekaterina Nguyen[1,2], Julien Barrier[1], Kornelia Indykiewicz[2], Wendong Wang[2], Roman V. Gorbachev[2], Na Xin[2], Kenji Watanabe[5], Takashi Taniguchi[5], Lee Hague[2], Vladimir I. Fal'ko[1,2], Irina V. Grigorieva[1], Leonid A. Ponomarenko[6], Alexey I. Berdyugin[3,4], Andre K. Geim[1,2]

[1] Department of Physics and Astronomy, University of Manchester, Manchester, UK
[2] National Graphene Institute, University of Manchester, Manchester, UK
[3] Department of Materials Science and Engineering, National University of Singapore, Singapore, Singapore
[4] Department of Physics, National University of Singapore, Singapore, Singapore
[5] National Institute for Materials Science, Tsukuba, Japan
[6] Department of Physics, University of Lancaster, Lancaster LA1 4YW, UK

**The electronic quality of two-dimensional systems is crucial when exploring quantum transport phenomena. In semiconductor heterostructures, decades of optimization have yielded record-quality two-dimensional gases with transport and quantum mobilities reaching close to $10^8$ and $10^6$ cm$^2$ V$^{-1}$ s$^{-1}$, respectively. Although the quality of graphene devices has also been improving, it remains comparatively lower. Here we report a transformative improvement in the electronic quality of graphene by employing graphite gates placed in its immediate proximity, at 1 nm separation. The resulting screening reduces charge inhomogeneity by two orders of magnitude, bringing it down to a few $10^7$ cm$^{-2}$ and limiting potential fluctuations to less than 1 meV. Quantum mobilities reach $10^7$ cm$^2$ V$^{-1}$ s$^{-1}$, surpassing those in the highest-quality semiconductor heterostructures by an order of magnitude, and the transport mobilities match their record. This quality enables Shubnikov–de Haas oscillations in fields as low as 1 mT, quantum Hall plateaus below 5 mT, and 10-µm-scale ballistic Dirac-fermion transport in the Boltzmann regime. Although proximity screening predictably suppresses electron-electron interactions, fractional quantum Hall states remain observable with their energy gaps reduced only by a factor of 3–5 compared to unscreened devices, demonstrating that many-body phenomena at spatial scales shorter than 10 nm remain robust. Our results offer a reliable route to improving a new level of electronic quality, which can be particularly valuable for studying more complex graphene and other two-dimensional systems where richer band structures and stronger interactions may reveal new physics enabled by reduced disorder.**

Among 2D systems, those based on GaAlAs heterostructures have continuously maintained the record for electronic quality[1–10]. Their latest milestone – transport mobilities $\mu$ up to ~5.7×10$^7$ cm$^2$ V$^{-1}$ s$^{-1}$ at sub-K temperatures ($T$) and electron densities of ~1.5×10$^{11}$ cm$^{-2}$ – came after three decades of painstaking work that yielded a 5-fold improvement in mobility[9,10]. Graphene, despite being a relatively recent addition to 2D systems, has also advanced in three big leaps. Starting from $\mu \approx 10^4$ cm$^2$ V$^{-1}$ s$^{-1}$ in early devices on oxidized Si wafers[11], its mobility first rose to ~10$^5$ cm$^2$ V$^{-1}$ s$^{-1}$ and then to ~10$^6$ cm$^2$ V$^{-1}$ s$^{-1}$ which was demonstrated for graphene both suspended[12,13] and encapsulated in hexagonal boron nitride (hBN)[14-17]. These mobilities have been observed for carrier densities $n \approx 10^{10}$–$10^{12}$ cm$^{-2}$ at liquid-helium $T$. While graphene holds the room-$T$ mobility record among all known materials ($\mu$ exceeds 0.15×10$^6$ cm$^2$ V$^{-1}$ s$^{-1}$ at $n \approx 10^{11}$ cm$^{-2}$, limited by phonon scattering)[17,18], its mobility at low $T$ is currently constrained by two extrinsic factors: edge scattering and charge inhomogeneity. Indeed, hBN-encapsulated graphene devices are typically less than 10 µm in size, making edge scattering dominant at high densities, whereas charge inhomogeneity $\delta n$ (exceeding a few 10$^9$ cm$^{-2}$ in best-quality devices) limits transport at low $n$. A recent study[17] using devices up to 40 µm wide showed that charged impurities also restricted $\mu$ at high $n$ to a few 10$^6$ cm$^2$ V$^{-1}$ s$^{-1}$. Another key measure of electronic quality is quantum mobility $\mu_q$ which determines the onset of Landau quantization and related phenomena such as Shubnikov – de Haas (SdH) oscillations and the quantum Hall effect (QHE). Highest-quality semiconductor heterostructures exhibit SdH oscillations in magnetic fields $B$ down to ~10 mT (refs. 9,19,20), which translates into $\mu_q \approx 10^6$ cm$^2$ V$^{-1}$ s$^{-1}$. For state-of-the-art graphene devices, it requires several times higher $B$ to observe SdH oscillations. Below we show that proximity gates[21] enable unprecedented quality: they allow $\mu_q \approx 10^7$ cm$^2$ V$^{-1}$ s$^{-1}$ and reduce inhomogeneity to ~3×10$^7$ cm$^{-2}$, that is, ~100 times lower than in the best devices with remote graphite gates and equivalent to one charge carrier left at the neutrality point (NP) per 10$^8$ carbon atoms. At low $n$, where edge scattering doesn't limit the mean free path $\ell$, transport mobilities reach 10$^8$ cm$^2$ V$^{-1}$ s$^{-1}$.

**Proximity-gated devices and their extraordinary homogeneity**

Our devices were double-gated multiterminal Hall bars fabricated from graphene monolayers sandwiched between two hBN crystals. The top hBN (20-70 nm thick) served as a gate dielectric for an evaporated Au/Cr electrode, whereas a graphite crystal was used as the bottom gate (inset of Fig. 1a); for details of fabrication, see Supplementary Information (SI). A distinguishing feature of our devices was ultrathin bottom hBN (3–4 atomic layers). Such a small thickness $d \approx 1$ nm was chosen to reduce the background electrostatic potential through image-charge screening, thereby suppressing electron-hole puddles and scattering. Indeed, in the presence of a metal gate, the background potential should be reduced proportionally to $[1 - \exp(-2\pi\alpha d/\mathcal{L})]$ where $\mathcal{L}$ is the characteristic size of the external potential variations, and $\alpha \approx 1.5$ accounts for the anisotropy of hBN's dielectric constant[21]. Given the large value of $2\pi\alpha$, potential variations larger in size than ~10 nm should be strongly suppressed for $d = 1$ nm. We could not employ thinner hBN because quantum tunneling caused considerable electrical leakage. Even with trilayer hBN, to avoid leakage when varying $n$, we could only use the top gate. The use of atomically flat graphite crystals as proximity gates was also essential to prevent trapped interfacial charges and electric-field fluctuations caused by surface roughness ('Device fabrication' in SI). Applied gate voltage was converted into $n$ using Hall measurements, which was essential because of a large contribution of quantum capacitance for small $d$ ('Converting gate voltages into carrier density' in SI). The main challenge in fabricating the described devices was to obtain sufficiently large (several hundred μm$^2$) few-layer hBN crystals, which limited the width $W$ of our Hall bars and, consequently, enhanced the role of edge scattering. We studied 7 such devices with $W$ from 6 to 10 μm (inset of Fig. 1b). For comparison, reference devices were also made, using the same procedures but with remote graphite gates ($d \gtrsim 20$ nm) and standard Si-wafer gates.

The impact of proximity screening on electronic quality is evident from Fig. 1a, which compares longitudinal resistivity $\rho_{xx}$ measured for devices with proximity and remote graphite gates. The reference device exhibited $\mu \approx 10^6$ cm$^2$ V$^{-1}$ s$^{-1}$ and the peak in $\rho_{xx}(n)$ narrower than 10$^{10}$ cm$^{-2}$, characteristics rarely achievable without the use of graphite gates. Remarkably, our proximity-gated devices exhibited much sharper peaks (Fig. 1a), indicating even higher mobility and better homogeneity. This observation qualitatively agrees with the previous reports where placing graphene monolayers directly on graphite or graphene resulted in improvement of optical and tunneling spectra[22,23]. To quantify charge homogeneity in our devices, we use the standard approach[24,25] and plot $\rho_{xx}(n)$ on a log-log scale (inset of Fig. 1a). At low $T$, graphene starts responding to gate doping only above a characteristic carrier density $\delta n$ because of electron-hole puddles. While the reference device exhibited $\delta n \approx 2 \times 10^9$ cm$^{-2}$ (Fig. 1 & Supplementary Fig. 2), matching the lowest $\delta n$ reported in the literature, proximity screening reduced $\delta n$ by as much as ~100 times (Fig. 1 & Supplementary Fig. 3). With increasing $T$, $\delta n$ increased (Fig. 1b) because of thermally excited electrons and holes, which appear at the NP in densities $n_{th} = (\pi/6)(k_B T)^2/(\hbar v_F)^2$ where $k_B$ and $\hbar$ are the Boltzmann and reduced Planck constants, and $v_F$ is graphene's Fermi velocity. The extra carriers broaden the $\rho_{xx}(n)$ peak, with their contribution described[18,25] by $\delta n \approx n_{th}/2$, in agreement with our measurements on proximity-gated devices (Fig. 1b & Supplementary Fig. 3b). For reference devices, residual inhomogeneity meant that $\delta n$ converged with the theoretical dependence only > 150 K (Supplementary Fig. 2c).

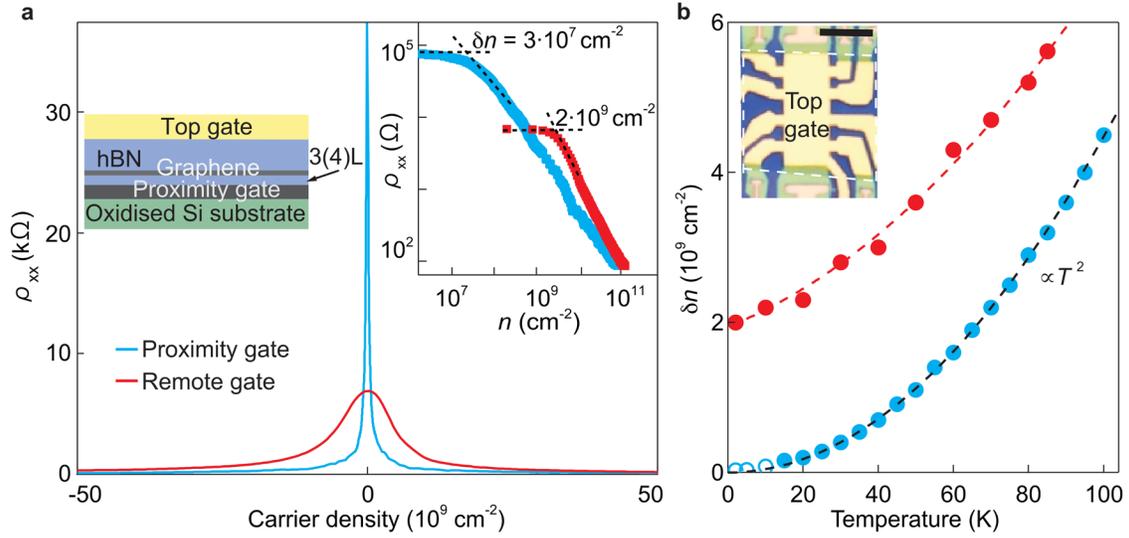

**Fig. 1| Profound effect of proximity screening on charge homogeneity. a** $\rho_{xx}(n)$ characteristic of state-of-the-art devices with remote graphite gates (red curve) and our proximity-gated devices (blue, device S1); $B = 0$, $T \approx 2$ K. Although the curves might look similar to many in the literature, the blue one is ~100 times narrower than for any previously reported device. The blue curve reaches ~100 k$\Omega$ but is cut off for clarity. Left inset: Schematic of proximity-gated devices. Right inset explains how $\delta n$ was evaluated. **b** Temperature dependence of $\delta n$ for the devices with proximity and remote gates (color coding as in **a**). Black parabola: $\delta n$ expected for perfect graphene. Red curve: Expected combined effect of residual inhomogeneity and thermal excitations[18]. Open blue circles indicate the low-$T$ regime affected by metal-insulator transition discussed in SI. Inset: Optical micrograph of one of our proximity-gated devices; scale bar, 10 μm. White dashed lines outline the graphite gate underneath.

Note that, below 10 K, proximity-gated devices often exhibited a logarithmic-kind increase in $\rho_{xx}(T)$ at the NP, with values reaching well above the resistance quantum $h/e^2$, indicating the emergence of a low-$T$ insulating state that could be destroyed by small $B \approx 1$ mT ('Metal-insulator transition' in SI). As this insulating state may affect our evaluation of $\delta n$ below 10 K, if using the standard procedure, we verified the values obtained in the low-$T$ regime using measurements of $\rho_{xy}(n)$. Hall resistivity switched sharply between electron and hole doping (Supplementary Fig. 3; 'Charge inhomogeneity from Hall measurements' in SI), and the width of the switching region provides an alternative quantitative measure of charge inhomogeneity[18], while $\rho_{xy}$ is expected to be less sensitive to localization effects. We found close agreement between $\delta n$ extracted using both Hall and longitudinal resistivities over the entire $T$ range, corroborating the 3-fold decrease in $\delta n$ between 2 and 10 K (Fig. 1c).

**Transport mobilities and ballistic Dirac plasma**

To evaluate $\mu$, we first employ the standard expression $\mu = 1/ne\rho_{xx}$. Its use necessitates a cutoff at a finite $n$ of a few $\delta n$, where electron-hole puddles start contributing to transport[18]. For reference devices, this means the cutoff at ~$10^{10}$ cm$^{-2}$ and places a limit on maximum achievable mobility as $\mu = (e\ell/\hbar)(\pi n)^{-1/2} \lesssim 10^7$ cm$^2$ V$^{-1}$ s$^{-1}$, assuming edge scattering dominates ($\ell \approx W \approx 10$ μm). At these $n$, our limited-size proximity-gated devices also cannot exceed this limit. However, their exceptional homogeneity pushes the cutoff to much lower $n$. Fig. 2a shows that $\rho_{xx}$ continues to scale approximately as $n^{-1/2}$ into the densities below $10^{10}$ cm$^{-2}$. For $n \gtrsim 3$–$5 \times 10^9$ cm$^{-2}$, extracted $\ell$ closely matches $W$ (red curve in Fig. 2a), as expected for ballistic transport. This also yields $\mu \approx 10^7$ cm$^2$ V$^{-1}$ s$^{-1}$.

For even lower $n$, but well outside the electron-hole puddle regime emerging below $10^8$ cm$^{-2}$, the mobility evaluated from $\rho_{xx}(n)$ continues to increase with decreasing $n$, reaching above $2.5 \times 10^7$ cm$^2$ V$^{-1}$ s$^{-1}$ at $n \approx 10^9$ cm$^{-2}$, a record for graphene to our knowledge. However, at such low $n$, the standard approach to evaluate $\mu$ can be questioned because extracted $\ell$ are comparable or start exceeding $W$ (Fig. 2a). While specular reflection at graphene edges can lead to $\ell > W$ (feasible because the Fermi wavelength $\lambda_F$ exceeds 1 µm), the estimate may be skewed by interference ('mesoscopic') fluctuations (see Fig. 2a and SI).

As an alternative means to determine $\ell$ and therefore $\mu$, we employed measurement geometries that can probe ballistic transport directly. One of them is magnetic focusing, where charge carriers injected from one contact are collected by another contact at distance $L$ (inset of Fig. 2b). Magnetic field bends trajectories, creating caustics that lead to focusing resonances[26,27]. Their positions matched accurately those expected theoretically (Fig. 2b & Supplementary Fig. 5). Importantly, the focusing resonances in proximity-gated devices survived down to $n \approx 10^9$ cm$^{-2}$ (Fig. 2b) whereas in our best reference device with $\mu \approx 7 \times 10^6$ cm$^2$ V$^{-1}$ s$^{-1}$, magnetic focusing could only be observed above $10^{11}$ cm$^{-2}$, and even higher $n$ were required for Si-gated devices (Supplementary Fig. 5). For the proximity-gated device in Fig. 2b, the observation of magneto-focusing peaks implies that charge carriers have $\ell \gtrsim \pi D_C/2 = \pi L/2 \approx 21$ µm ($D_C$ is the cyclotron diameter). Using the semi-quantitative criterion detailed in SI ('Magnetic focusing in reference devices'), we find that the persistence of focusing peaks down to $10^9$ cm$^{-2}$ yields $\mu \gtrsim 6 \times 10^7$ cm$^2$ V$^{-1}$ s$^{-1}$, consistent with our estimates above from the $\rho_{xx}(n)$ dependence.

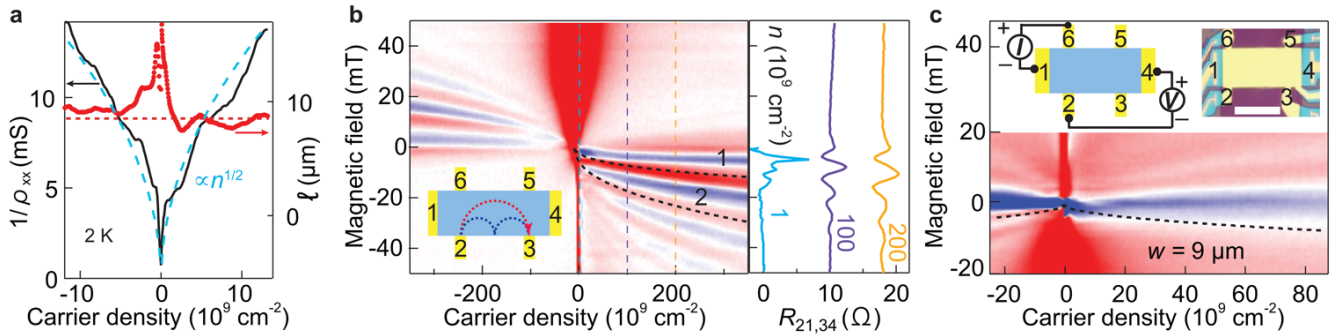

**Fig. 2| Ballistic transport in proximity-screened graphene. a** Conductivity and mean free path (black and red curves, respectively). Blue dashed curve: $n^{1/2}$ dependence expected for transport limited by edge scattering (best fit yields $\ell \approx 9$ µm). Red line indicates the actual device width of ~8.5 µm. **b** Ballistic transport probed by magnetic focusing. Left: Map of focusing resistance $R_{21,34}(n, B) = V_{34}/I_{21}$ (blue-to-red scale, -5 to 5 Ohm). Current $I_{21}$ is driven between contacts 2 and 1 as shown in the inset; voltage $V_{34}$ is measured between 3 and 4; $L \approx 13.5$ µm. Right: Vertical cuts at fixed $n$ marked in the map by color-coded dashed lines. Black dashed curves: Expected positions of the first two focusing peaks (corresponding trajectories shown in the inset). **c** Example of bend-resistance measurements with insets showing the measurement geometry (left) and the device's optical micrograph (right; scale bar, 10 µm). Color map shows $R_{61,42}$ (color scale as in **b**). Dashed curve: $W = D_C/2$, the condition where the bend resistance is expected to reverse its sign[15,26]. Data in **a** are for device S4 at 2 K; data in **b**, **c** are for device S6 (at 20 K to suppress mesoscopics).

Next, we employed the bend-resistance geometry[15,26] (Fig. 2c). In zero $B$, the measured bend resistance remained negative at all $n$, showing that carriers travelled ballistically between the injector and collector (that is, $\ell > W \approx 9$ µm). Even at the NP, where Dirac fermions formed a Boltzmann rather than Fermi gas and experienced frequent Planckian scattering[25] ('Metal-insulator transition' in SI), transport in proximity-gated devices remained ballistic. The carrier density in the Dirac plasma is given by $n_{th}$ and is ~$3 \times 10^8$ cm$^{-2}$ for the 20 K measurements in Fig. 2c. This yields $\mu = (e\ell/\hbar)(\pi n)^{-1/2} \geq 5 \times 10^7$ cm$^2$ V$^{-1}$ s$^{-1}$, in close agreement with the $\mu$ found above in the Fermi regime.

The negative bend resistance at the NP was observed down to 10 K, suggesting that, in the low-$T$ Dirac plasma, $\mu$

could exceed $10^8$ cm$^2$ V$^{-1}$ s$^{-1}$. However, interference fluctuations ($\lambda_F \gtrsim 3$ μm) emerging at these $T$ started crisscrossing the color map near zero $n$, making the last estimate less reliable. Independent evidence for ballistic transport in the Dirac plasma at $T$ close to 10 K comes from $\rho_{xx}(T)$ dependence at the NP (Supplementary Fig. 4). Above 50 K, proximity-gated devices exhibited the expected constant resistivity of ~1 kOhm, characteristic of Planckian scattering[25,28]. At lower $T$, but before the onset of the insulating state at ~10 K, $\rho_{xx}$ was found to evolve as $1/T$ (Supplementary Fig. 4c). The $T$ dependence yields scattering times shorter than the Planckian limit and given by $W/v_F$ ('Metal-insulator transition' in SI), thus confirming ballistic transport limited by edge scattering down to 10 K. This supports our above estimate for $\mu \gtrsim 10^8$ cm$^2$ V$^{-1}$ s$^{-1}$ in the low-$T$ Dirac plasma of proximity-gated devices.

Direct comparison of graphene with other 2D systems in terms of their transport mobilities is unfortunately hindered by very different density ranges where ballistic transport is observed. The record mobilities for GaAlAs heterostructures were achieved in a narrow range of $n \approx 1-2 \times 10^{11}$ cm$^{-2}$ whereas, at such densities, mobility even in our reference devices was limited by their $W$. Additionally, the scattering mechanisms fundamentally differ between the two systems. Transport in high-quality hBN-encapsulated graphene is primarily limited by edge scattering and background potential fluctuations ('Sources of disorder' in SI) whereas mobilities in GaAlAs heterostructures are limited by ionized impurities[6,8,9]. The different electronic spectra (linear vs parabolic dispersion) also affect scattering efficiency. Nonetheless, as $\mu$ generally increases with $n$ due to improved self-screening[17], even higher mobilities are probably achievable in graphene at the densities typical of GaAlAs heterostructures, if larger proximity-gated devices become available to reduce edge scattering.

**Record quantum mobilities**

A more direct comparison between graphene and other 2D systems can be made using quantum mobility that is insensitive to device size and geometry. Our proximity-gated devices demonstrated Landau quantization in strikingly low $B$ (Fig. 3a). For example, at 50 mT and 2 K, we observed more than 15 SdH oscillations (filling factor $\nu > 60$; cyclotron gaps between neighboring levels down to 1.1 meV). This indicates extremely narrow Landau levels, in excellent agreement with the measured $\delta n$ that yielded Dirac point's spatial variations of $\hbar v_F (\pi \delta n)^{1/2} \approx 0.5$ meV. Furthermore, SdH oscillations became apparent at $B^* \approx 1-2$ mT (Fig. 3b, Supplementary Fig. 6), much lower than the onset fields $B^*$ in reference devices with remote graphite and Si-wafer gates ($\gtrsim 35$ and 80 mT, respectively). To ensure that the observed minima in $\rho_{xx}(n)$ originated from Landau quantization rather than interference fluctuations, we checked not only for their correct positions in the $n$-$B$ parameter space (Fig. 3b) but also shifted mesoscopic features by cycling up to room $T$ and varying the electric field across graphene using both top and bottom gates. These procedures altered both the rapidly rising background near the NP and fluctuations' positions but left SdH oscillations unaffected, confirming that the 1-mT minima were indeed due to Landau quantization. We believe that SdH oscillations in proximity-gated devices appeared at even lower $B$ but were obscured until their amplitude became comparable to $\rho_{xx}$ (background subtraction reveals features down to ~0.5 mT). Note that with a constant background, SdH oscillations typically become visible at amplitudes of just several % of $\rho_{xx}$ ('Onset of SdH oscillations' in SI). The use of higher $T$ to suppress mesoscopic fluctuations was unpractical as it smeared cyclotron gaps (1.2 meV for $\nu = \pm 2$ at 1 mT), while reducing $T$ below 2 K amplified the fluctuations (Supplementary Fig. 4a).

Quantum mobility can be estimated from the onset of SdH oscillations using expression $\mu_q B^* \approx 1$. Although this criterion is widely used in the literature, we validated its quantitative accuracy through studying SdH oscillations as a function of $B$ (Supplementary Fig. 7). Their amplitude followed the Dingle formula $\exp(-\pi/\mu_q B)$, and we found $\mu_q B^* = 1$ to hold within 10–20%, if the first discernible oscillation was considered as $B^*$ ('Determination of quantum mobility' in SI). Thus, the Landau quantization visible on our plots at $B = 1$ mT yields $\mu_q$ of at least $10^7$ cm$^2$ V$^{-1}$ s$^{-1}$, setting a lower bound for quantum mobility in proximity-gated devices (as discussed in SI, magnetic fields larger than the actual $B^*$ were required to resolve oscillations against the rising background and mesoscopics). Since Landau quantization involves completed cyclotron orbits, this implies $\ell \geq \pi D_c$, even if small-angle scattering – that disrupts quantization but has a lesser effect on transport – is ignored. Using this $\ell$ in the expression for transport mobility gives $\mu \geq 2\pi\mu_q > 6\times10^7$ cm$^2$ V$^{-1}$ s$^{-1}$, consistent with our estimates from $\rho_{xx}(n)$.

In line with the early onset of SdH oscillations, the QHE fully developed at $B \leq 5$ mT (Fig. 3c & Supplementary Fig. 8a). The transition between electron and hole plateaus at the NP had a half-width at half-height of $\sim 3\times10^7$ cm$^{-2}$ (Fig. 3c), providing an independent measure of charge inhomogeneity in proximity-gated devices. Importantly, the emergence of QHE plateaus was not constrained by graphene's quality but depended on the width $w$ of voltage probes, with plateaus developing earlier in devices with larger $w$ (Supplementary Fig. 8b). This is attributed to the fact that, at such low $B$, both $D_c \approx 0.7$ μm at 5 mT and the magnetic length $\ell_B = D_c(\nu/8)^{1/2}$ were comparable to $w \leq 1$ μm, causing edge states to be partially reflected from the contacts and preventing the development of quantized plateaus[29].

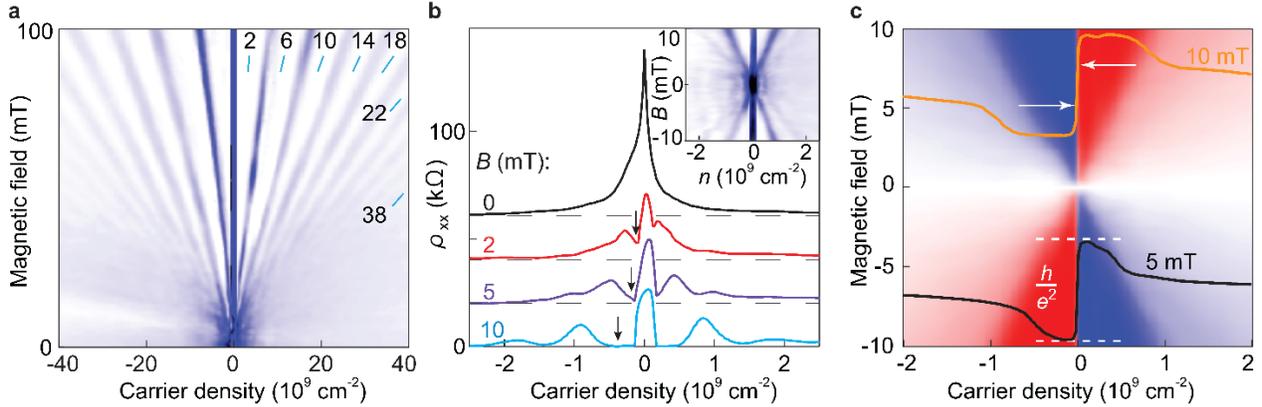

**Fig. 3| Quantization in milli-Tesla fields. a** Landau fan diagram $\rho_{xx}(n,B)$ (white-to-blue scale, 0 to 4 kOhm). Numbers with blue dashes denote $\nu$. **b** Horizontal cuts from **a** at different $B$. Inset: Details of the fan diagram in low $B$ (white-to-blue, 0 to 40 kOhm). Arrows: Expected positions for $\nu = -2$. Note that $\rho_{xx}(n)$ changes rapidly near the NP, which results in a wide dark region that obscures the onset of SdH oscillations. They are better resolved on the horizontal cuts (also, see Supplementary Fig. 6). **c** Map for $\rho_{xy}$ (blue-to-red scale, $\pm\frac{1}{2}$ $h/e^2$). Superimposed curves: $\rho_{xy}(n)$ traces at 5 and 10 mT (offset for clarity). Arrows mark the full transition width at half-height, $\sim 6\times10^7$ cm$^{-2}$. All data are for device S1 at $\sim 2$ K.

**Effect of proximity screening on many-body phenomena**

The achieved electronic quality comes with a tradeoff. Many-body phenomena – which are of considerable interest and tend to emerge with each order-of-magnitude improvement in quality – are unavoidably suppressed by enhanced screening[21,30]. To assess how proximity screening affects many-body phenomena in graphene, we studied the fractional QHE in our devices ($B$ up to 12 T and $T$ down to 50 mK). Figure 4a shows distinct Hall plateaus and the corresponding resistivity minima at $\nu = 2/3, 5/3, 8/3, 10/3$ and $11/3$. Notably, no signatures of the 1/3 state could be observed, despite it usually being most profound. We attribute its absence to negative quantum capacitance that can suppress this state in gate voltage measurements[31] ('Fractional quantum Hall effect under proximity screening' in SI). We extracted fractional QHE energy gaps (Fig. 4b) and compared them with those in our reference devices that exhibited gaps consistent with those reported in literature[32-34] (Fig. 4c). While proximity screening reduced the fractional gaps by a factor of 3-5 compared to unscreened devices, the gaps remained substantial, larger than those typical of other 2D systems.

The suppression factor can be understood as follows. The interaction energy responsible for the fractional QHE scales as $\propto e^2/\varepsilon \ell_B$ (two electrons at distance $\ell_B$ signifying the spatial scale of electron-electron interactions in magnetic field and $\varepsilon$ is the effective dielectric constant) whereas proximity screening changes this into $\propto e^2 2d/\varepsilon \ell_B^2$ (two interacting dipoles made of an electron and its image charge at distance $2d$)[30]. This yields a suppression factor of $\ell_B/2d \approx 4$ at 12 T, in good agreement with our observations (Fig. 4c). The fractional states first appeared above 7 T and became more pronounced at higher fields where $\ell_B$ decreased below 10 nm. This length scale matches well the spatial range over which the proximity screening with $d \approx 1$ nm suppresses background electrostatic potential because of the discussed large factor $2\pi\alpha$. The same 10 nm scale was previously noticed in suppression of two other types of interaction phenomena (electron viscosity and Umklapp scattering)[21]. These observations suggest that while proximity screening does reduce interaction strengths, many-body physics involving scales shorter than 10 nm should remain accessible in proximity-gated devices.

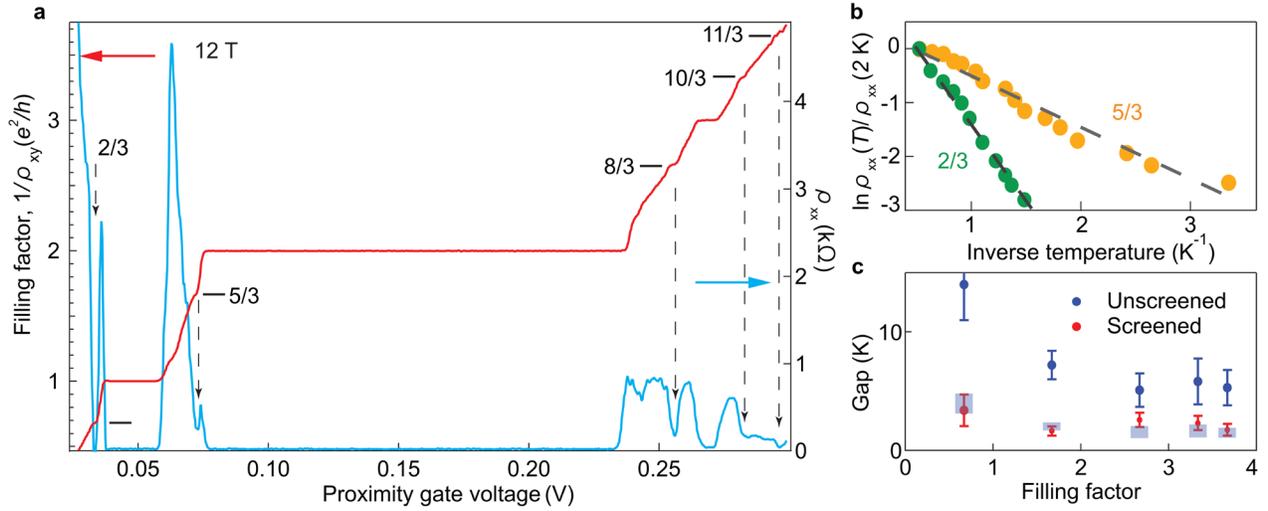

**Fig. 4| Fractional QHE in proximity-gated devices. a** $\rho_{xx}$ and $\rho_{xy}$ at 12 T and 50 mK (blue and red curves; left and right axes, respectively). Data are plotted as a function of proximity gate voltage, as accurate conversion to carrier density was unfeasible due to the 2.5D QHE in the graphite gate[35] and negative quantum capacitance[31]. $\rho_{xy}$ is plotted in terms of $\nu = (h/e^2)/\rho_{xy}$. Horizontal lines mark expected positions of fractional plateaus, while arrows indicate the corresponding $\rho_{xx}$ minima. **b** Arrhenius plots for resistance minima (normalized by values at 2 K) at $\nu$ = 2/3 and 5/3 were used to extract activation energies. **c** Comparison of fractional QHE gaps in proximity-gated device S1 (red symbols) and a remote-gate device (blue symbols with error bars). Blue rectangular symbols: Expected gaps after proximity screening, calculated using the $\ell_B/2d$ suppression factor with $d$ = 1 nm and $\ell_B \approx 7.5$ nm for 12 T.

**Conclusion**

Our study demonstrates that proximity screening can improve electronic quality of graphene by up to two orders of magnitude. The resulting charge homogeneity is unprecedented (Dirac point fluctuations of less than 10 K), enabling extremely narrow Landau levels and the QHE in a few mT. While the achieved quality comes at the expense of suppressing many-body phenomena, interactions involving relatively short spatial scales (< 10 nm) remain strong, suggesting that proximity screening may be particularly valuable for studying short-range correlated states and many-body physics in high magnetic fields. We anticipate the approach will prove especially beneficial for studying graphene multilayers and superlattices. As the quality of 2D semiconductors continues to improve, proximity screening may also be relevant for these systems. Alternatively, our approach can be used to intentionally suppress many-body interactions while simultaneously delivering superior electronic quality, as demonstrated by the observation of the helical QHE[36] in our proximity-gated devices at fields below 80 mT (SI).

# Supplementary Information

## #1 Device fabrication

The devices were fabricated using the standard van der Waals assembly and electron-beam lithography, following procedures established in the literature without introducing critical modifications. Our protocols are summarized below.

Monolayer graphene, hBN crystals (both 30-70 nm thick and 3-4 layers thin) and relatively thick graphite (5-50 nm) were mechanically exfoliated onto an oxidized Si wafer (290 or 70 nm $SiO_2$). Their thicknesses were determined using optical contrast, atomic force microscopy and Raman spectroscopy. For most of the reference devices with remote graphite and Si gates, we assembled the exfoliated crystals using polymer-free flexible silicon nitride membranes[17]. For all proximity-gated devices and some reference devices, we employed polydimethylsiloxane stamps with polypropylene carbonate as a sacrificial layer. While not critical, we found the latter method preferable for proximity-gated devices as it imposes fewer constraints.

The assembly of encapsulated graphene began by picking up a selected top hBN crystal and then using it to sequentially collect graphene and bottom hBN crystals. Crucially for achieving ultrahigh electronic quality, the stacking process required slow deposition of 2D crystals onto each other, which minimized bubble and wrinkle formation and consequently allowed the sufficiently large final devices (Supplementary Fig. 1). All assembly procedures were conducted in air. Exfoliated flakes were either used within an hour or stored in a glove box for assembly within 2-3 weeks. The completed trilayer stacks were transferred onto either a graphite crystal (for graphite-gated devices) or an oxidized Si wafer.

Following the assembly, we used electron-beam lithography to define the top gate region and deposited Cr-Au metallization by electron-beam evaporation. In the subsequent lithography step, we created electrical contacts by first exposing graphene edges using reactive-ion etching and then depositing Cr-Au to form one-dimensional contacts. The typical contact resistance was 2-5 kΩ·μm at the NP and decreased down to 0.1-0.3 kΩ·μm at high doping. In the final step, we used the metallic top gate as an etching mask and defined multiterminal Hall bars, as shown in the micrographs throughout the main text and Supplementary Information.

Although some devices failed during fabrication (typically because of poor 1D contacts), all successfully fabricated proximity-gated devices demonstrated consistently ultrahigh quality, with SdH oscillations always emerging below 4-5 mT and sometimes below 1 mT. The primary determinants of our device quality were interface cleanliness and careful, slow stacking (as described above).

## #2 Electrical measurements

Magnetotransport measurements were carried out in two cryogenic systems. A liquid-He cryostat with a variable temperature insert was used for studying transport from 1.7 K up to room $T$. For measurements in the fractional QHE regime, we used a dilution refrigerator (*Oxford Instruments*) with $T$ from 6 K down to 50 mK and magnetic fields up to 12 T.

For resistance measurements, we used the standard low-frequency lock-in technique at 30.5 Hz. Typically, ac currents were between 1 and 5 nA, which allowed us to avoid heating and self-gating effects. Higher currents were required in magneto-focusing and bend-resistance experiments to achieve sufficient signal-to-noise ratios: up to 1 μA for devices with remote gates and up to 0.1 μA for proximity-gated devices.

To control carrier density $n$, we applied gate voltages using *Keithley Sourcemeter-2636B*. For proximity-gated devices, in most cases we used only the top Au gate while keeping the bottom graphite or Si gate grounded. This approach prevented tunneling through and possible breakdown of the few-layer hBN dielectric between graphene and the proximity gate. However, for measurements of the fractional QHE, we found it advantageous to use the proximity gate to vary *n*. This strategy allowed us to avoid insulating QHE states inside the graphite gate (due to the 2.5-dimensional QHE in graphite[35]). Simultaneously, we applied a large bias to the bottom Si gate to suppress insulating states in regions of graphene's electrical leads that were not covered by the graphite gate.

**#3 Converting gate voltages into carrier density**

For devices with relatively thick gate dielectrics, carrier density *n* in graphene exhibits a practically linear dependence on gate voltage $V_g$ because the geometric capacitance dominates (~12 nF/cm² for our Si-gated devices). The linearity holds due to typical shifts in graphene's chemical potential being small compared with $eV_g$. However, the simple relationship $n \propto V_g$ breaks down dramatically in proximity-gated devices with their ultrathin dielectrics because high-quality graphene exhibits a vanishingly small density of states (DoS) near the NP and in quantizing magnetic fields. This phenomenon is conventionally described in terms of quantum capacitance $C_q = e\partial n/\partial \mu$, which contributes in series with the geometric capacitance and consequently dominates if the DoS and hence $C_q$ become sufficiently small[37,38]. The resulting nonlinearities in the $n(V_g)$ dependence become particularly pronounced in ultrahigh quality graphene, as elucidated by Supplementary Fig. 1. Below we describe our methodology for converting gate voltage into carrier density.

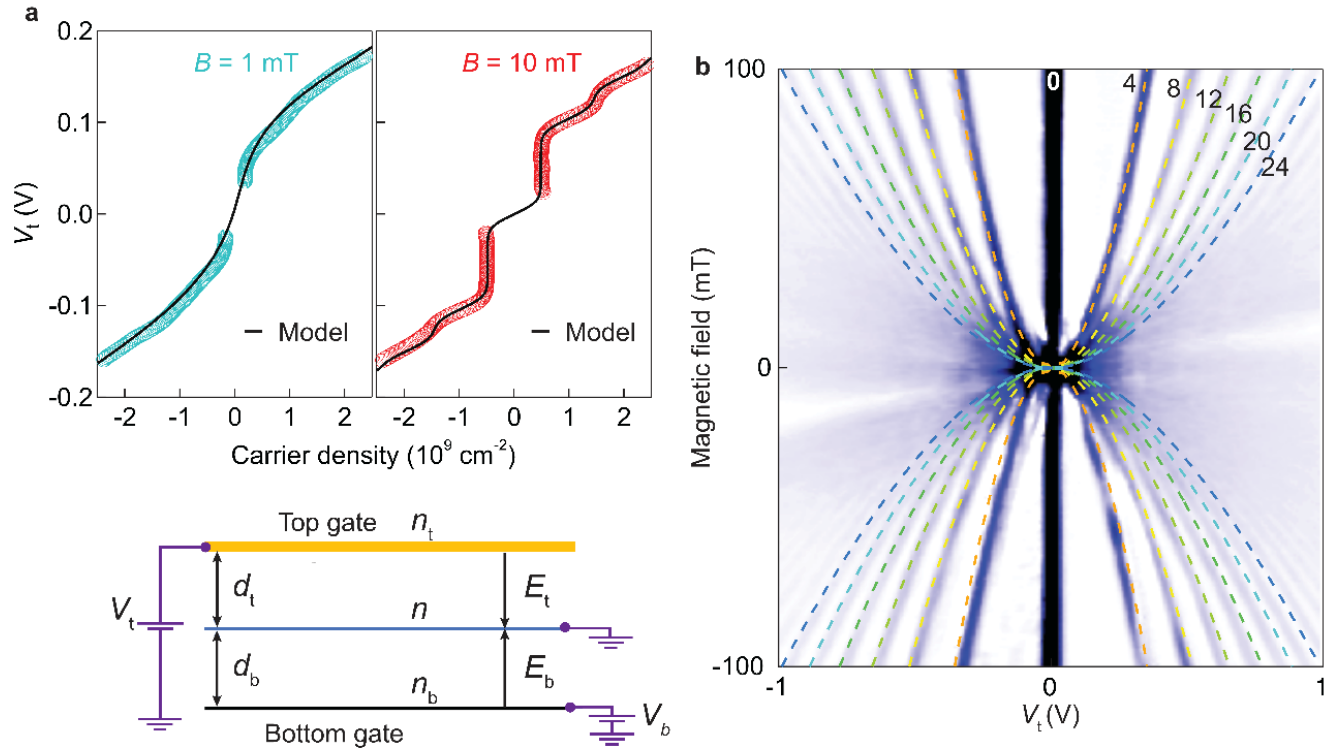

**Supplementary Fig. 1| Converting gate voltages into carrier density**. **a** Gate voltage $V_t$ as a function of carrier density *n* determined from $\rho_{xy}$ measurements at 1 and 10 mT (left and right panel, respectively). Symbols: Experimental data. Solid curves: Best fits using our quantum capacitance model with Landau level broadening $\Gamma$ = 0.25 meV. Bottom panel: Device schematics. Applied voltages $V_t$ and $V_b$ create electric fields $E_t$ and $E_b$ in hBN layers (thicknesses $d_t$ and $d_b$). Charge densities $n_t$ and $n_b$ at the gates induce the carrier density *n* in graphene. **b** Landau fan diagram $\rho_{xx}(V_t,B)$ measured at 2 K for proximity-gated device S1 (white-to-blue scale: 0 to 4 kOhm). Dashed curves: Calculated positions of peaks in $\rho_{xx}$ for the labeled filling factors.

Our devices can be modelled as parallel-plate capacitors with top and bottom gates biased at $V_t$ and $V_b$ (bottom panel of Supplementary Fig. 1a). The gates are separated from graphene by hBN layers of thicknesses $d_b$ and $d_t$. Applied voltages create electric fields $E_t$ and $E_b$ inside the top and bottom hBN dielectrics. $V_t$ shifts the electrostatic potential by $d_t E_t$ and therefore changes graphene's electrochemical potential $\mu(n)$, yielding the equation $eV_t = ed_t E_t + \mu(n)$ where $e$ denotes the absolute value of the electron charge. Similarly for the bottom gate, $eV_b = ed_b E_b + \mu(n)$. The charge densities at gate interfaces are $n_t = -\varepsilon\varepsilon_0 E_t/e$ and $n_b = -\varepsilon\varepsilon_0 E_b/e$ where $\varepsilon_0$ is the vacuum permittivity and $\varepsilon \approx 3.5$ is the hBN dielectric constant. Charge conservation requires $n_t + n_b + n = 0$, which yields the following equation connecting $n$ and $\mu$ with the gate voltages

$$V_b + \frac{d_b}{d_t} V_t = \frac{ed_b}{\varepsilon\varepsilon_0} n + \left(1 + \frac{d_b}{d_t}\right) \frac{\mu(n)}{e} \tag{1}$$

This equation is generally applicable to any double-gated field-effect device that uses a 2D material as a conducting channel. In most of our measurements, we used only the top gate to vary $n$ while keeping both graphene and the proximity (bottom) gate grounded (that is, $V_b = 0$). It is important to note that, according to the above equations, electric fields still develop in the gate dielectrics even when graphene is electrically connected to one of the gates, provided the other gate shifts $\mu(n)$ from zero. Indeed, if $V_b = 0$, we obtain $E_b = -\mu(n)/ed_b$ which shows that, despite being grounded, the proximity gate with its ultrathin dielectric actively participates in the device electrostatics and, as $E_b \propto 1/d_b$, significantly contributes to nonlinear behavior of $n(V_t)$. The electrochemical potential $\mu(n)$ in Eq. 1 can be calculated by numerically inverting the integral

$$n(\mu) = \int_0^\infty D(E) f(E, \mu, T) dE$$

where $f(E,\mu,T)$ is the Fermi-Dirac distribution function, and the DoS in graphene $D(E)$ is modelled as

$$D(E) = \frac{4eB}{\sqrt{2\pi}\hbar\Gamma} \sum_i \exp\left(-\frac{(E-E_i)^2}{2\Gamma^2}\right)$$

This equation is written for the general case of a finite magnetic field. Here, $E_i$ is the energy of the $i$-th Landau level, and $D(E)$ depends not only on energy $E$ but also on magnetic field $B$ and level broadening $\Gamma$ due to both temperature and disorder[37,38]. We solved Eq. 1 numerically (using an energy cutoff of 0.5 eV in all computations). Examples for non-quantizing and quantizing fields are illustrated in Supplementary Fig. 1a by black solid curves. Experimentally, we determined the carrier density $n$ as a function of $V_t$ by measuring the Hall resistivity $\rho_{xy}(V_t)$ at zero $V_b$ and employing the standard expression $n = B/e\rho_{xy}$. The latter is valid only outside the mixed electron-hole regime near the NP (see, for example, Supplementary Fig. 3a and section 'Charge inhomogeneity from Hall measurements'), which resulted in gaps on our experimental curves for very low $n \lesssim \delta n$ (Supplementary Fig. 1a). For the known thicknesses $d_b$ and $d_t$, we fitted our experimental data using the numerical model described above with broadening $\Gamma$ as the only fitting parameter. Supplementary Fig. 1a shows good agreement between the experimental $n(V_t)$ dependences and the model, yielding $\Gamma \approx 0.25$ meV. This shows that, for the presented device, the energy broadening was dominated primarily by $T$. Our model accurately described the experimental curves in both non-quantizing and quantizing fields (left and right panels, respectively). We applied these fitted dependences (using the $\Gamma$ values determined for each device) to convert $V_t$ into $n$ and present the experimental data for proximity-gated devices as a function of $n$ for different $B$ and $T$.

The accuracy of our conversion procedure can be assessed by examining how well it describes the $\rho_{xx}$ peak positions on the Landau fan diagram measured over a wide range of $V_t$ and $B$ (Supplementary Fig. 1b). For monolayer graphene in moderate $B$, these peaks should appear at filling factors $v = 0, \pm4, \pm8, \pm12, \ldots$ and follow the immutable relation $B(v) = nh/ev$ on the fan diagram. Using our model with the extracted $\Gamma$ values, we calculated $n(V_t)$ and used this to determine the $B(V_t)$ dependences for different filling factors, as shown by dashed curves in Supplementary Fig. 1b. The excellent agreement between measured and predicted Landau level positions confirms the accuracy of our procedure for converting gate voltages into carrier densities. Using the same procedure, we transformed the nonlinear fan diagrams obtained as functions of $V_t$ (such as shown in Supplementary Fig. 1b) into the standard fan diagrams where Landau levels appear as straight lines radiating from the origin, as required by theory (Fig. 3 and Supplementary Fig. 6). These linear fan diagrams provide additional confirmation of our approach's accuracy.

#### #4 Charge inhomogeneity in best remote-gated devices

In the main text, we compared our proximity-gated devices and reference devices that used remote graphite gates. Supplementary Fig. 2 provides additional details about our best remote-gated device. Its zero-$B$ resistivity $\rho_{xx}(n)$ near the NP is shown for two representative $T$ (Supplementary Fig. 2a) and replotted on a log-log scale to evaluate the residual charge inhomogeneity $\delta n$ (Supplementary Fig. 2b). With increasing $T$, the resistivity peak broadens because of thermally excited carriers. The extracted $\delta n(T)$ is shown in Supplementary Fig. 2c. Above 150 K, the experimental data follow the expected parabolic dependence $\delta n = \frac{1}{2}n_{th}(T) \propto T^2$ within our measurement accuracy, as discussed in the main text. Below 10 K, $\delta n$ saturates because electron-hole puddles dominate transport properties.

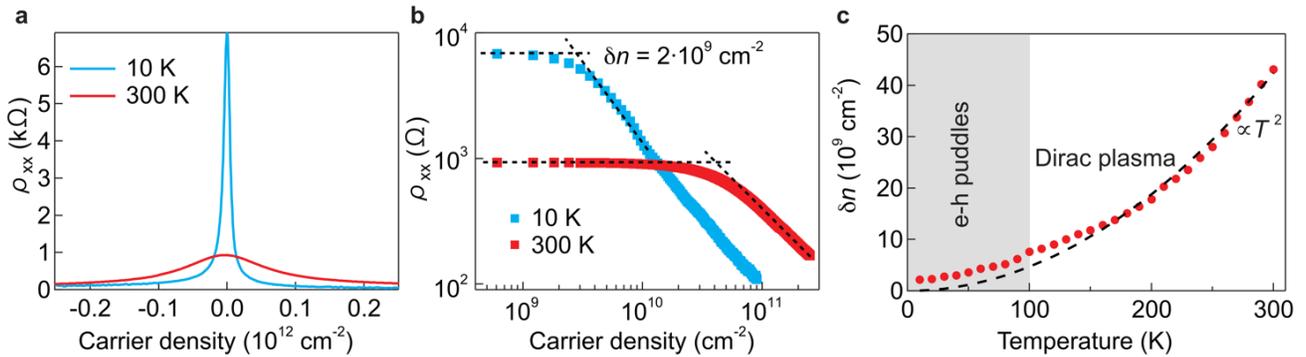

**Supplementary Fig. 2| Transport in hBN-encapsulated graphene with remote graphite gates. a** Zero-field $\rho_{xx}(n)$ at 10 and 300 K. **b** Same data on a log-log scale. Dashed lines show our evaluation of $\delta n$. **c** Temperature dependence of $\delta n$. Symbols: Experimental data. Black curve: Expected broadening due to thermally excited carriers. Grey region indicates that electron-hole puddles dominate. The region's upper bound corresponds to $T$ at which thermal broadening becomes twice that given by the puddles.

#### #5 Charge inhomogeneity from Hall measurements

An alternative way to evaluate charge inhomogeneity $\delta n$ is to use Hall measurements[18]. Supplementary Fig. 3a shows $\rho_{xy}(n)$ for one of our proximity-gated devices at two representative $T$. The density range between the extrema in $\rho_{xy}(n)$ marks the regime of mixed electron-hole transport where Hall response no longer follows the standard single-carrier dependence $\rho_{xy}(n) = eB/n$. The distance between the extrema is given by $n_{th} \approx 2\delta n$ and provides another quantitative measure of charge inhomogeneity[18]. With increasing $T$ from 10 to 40 K, the extrema move apart and broaden. Supplementary Fig. 3b compares the extracted $\delta n$ with that expected from thermal excitations at the NP. Above 10K, $\delta n$ follows the parabolic dependence $\tfrac{1}{2}n_{th}(T) \propto T^2$ (within our measurement accuracy), matching the results from the $\rho_{xx}(n)$ peak broadening in Fig. 1c of the main text.

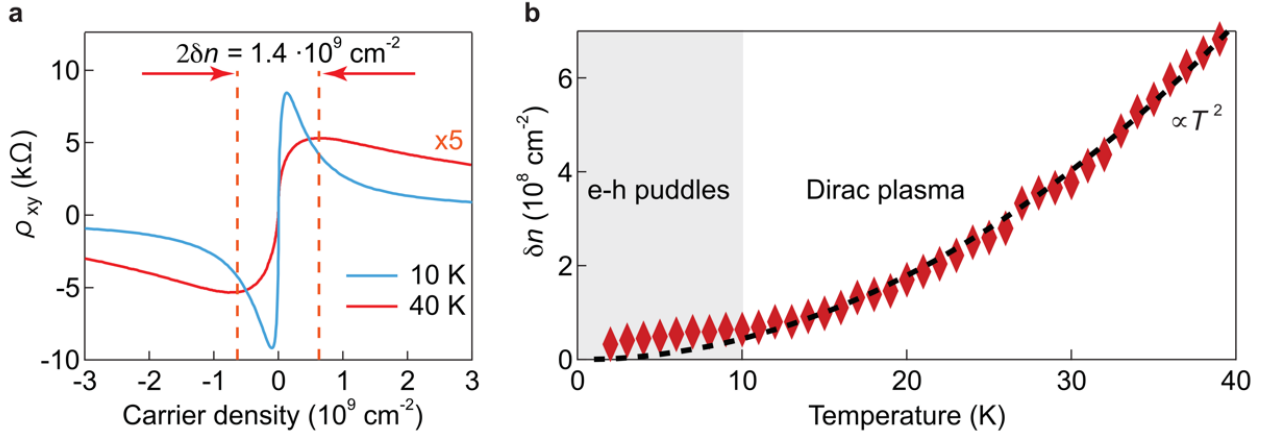

**Supplementary Fig. 3 | Estimating charge inhomogeneity using Hall resistivity. a** $\rho_{xy}(n)$ near the NP at 10 and 40 K (color-coded curves). Measurements at 3 mT provided sufficiently large Hall voltages for high accuracy. The full width between the dashed lines corresponds to $2\delta n$ for the case of 40 K. **b** Temperature dependence of $\delta n$ (symbols). Black dashed curves: $\tfrac{1}{2}n_{th}(T)$. Grey region indicates the $T$ range affected by electron-hole puddles.

**#6 Metal-insulator transition**

At low $T$, our proximity-gated devices exhibited a highly resistive state at the NP, which showed signatures of strong localization and quantum interference. Supplementary Fig. 4a plots $\rho_{xx}(n)$ in zero $B$, which displays a pronounced peak at the NP reaching ~100 kΩ at 0.5 K, that is, 5 times larger than expected for charge-neutral ballistic graphene in its minimum conductivity state[39]. This behavior indicates a metal-insulator transition, probably of Anderson type, as previously observed in double-layer graphene heterostructures[40]. The resistivity overshoots the maximum metallic value[39] of $\pi h/4e^2$ at $n \lesssim 10^8$ cm$^{-2}$. This corresponds to the Fermi wavelength $\lambda_F \gtrsim 3.5$ μm, which becomes comparable to the device width $W$. In this regime, the number of electronic channels fitting $W$ is reduced to less than 5, while the Fermi energy is only ~10-15 K. Under these conditions, quantum confinement effects become significant and open energy gaps between quantized subbands, which can become comparable to $k_BT$. The onset of spatial quantization could contribute to the observed insulating state, although Anderson-type localization is probably the dominant mechanism, given the observed $B$ and $T$ dependences[40]. Away from the NP, additional 'mesoscopic' features appeared on the $\rho_{xx}(n)$ curves at sub-K $T$ (Supplementary Fig. 4a). Given the long $\lambda_F$, these features can be attributed to interference (Fabry-Perot) resonances caused by standing waves in our ballistic devices.

To elucidate the nature of the observed insulating state, we analyzed the $T$ dependence of resistivity at the NP ($\rho_{NP}$). Supplementary Fig. 4c shows an example of the measured $\rho_{NP}(T)$, while the same data are replotted in the Arrhenius coordinates in Supplementary Fig. 4b. The complete evolution of $\rho_{xx}(n)$ with $T$ is presented in Supplementary Fig. 4d. These plots rule out the presence of simple thermal activation over an energy gap at the Dirac point, which would manifest as a straight line in the Arrhenius plot. Instead, $\rho_{NP}(T)$ reveals two distinct regimes: 1) logarithmic increase with decreasing $T$ below 10 K (black dashed curve in Supplementary Fig. 4c) and 2) $1/T$ dependence at higher temperatures (red curve). The $1/T$ behavior can be understood as follows: the density of thermally excited electrons and holes at the Dirac point is given by $n_{th} \propto T^2$ (see Fig. 1c of the main text), while their effective masses $m_{th} \propto T$ (refs. 18,25). Consequently, the observed dependence $\rho_{NP} \propto 1/T$ implies a temperature-independent scattering time, $\tau = m_{th}/(e^2 n_{th}\rho_{NP})$, consistent with Dirac fermions' ballistic transport limited by edge scattering and $\tau \approx W/v_F$, as discussed in the main text.

The magnetic field dependence provides further insight into the metal-insulator transition. Supplementary Fig. 4e shows $\rho_{NP}(B)$ at 2 K. The insulating state is suppressed by fields of ~1 mT, resulting in pronounced negative magnetoresistance. We attribute this suppression to breaking time-reversal symmetry for interfering electron trajectories, a mechanism analogous to the destruction of weak localization in normal metals and also expected to play a role in strong (Anderson-type) localization[18]. With increasing $T$ above 10 K, the magnetoresistance changes sign from negative to positive, as evident from comparing $\rho_{NP}(T)$ at 0 and 5 mT in Supplementary Fig. 4f. This positive magnetoresistance is characteristic of the charge-neutral Dirac plasma in the Boltzmann regime[25]. The sign change occurs at ~10 K, providing independent confirmation of the regime change from the insulating state to the ballistic Dirac plasma, which was inferred above from changes in the functional form of $\rho_{NP}(T)$ (Supplementary Fig. 4c). The negative magnetoresistance also rules out an excitonic gap at the Dirac point. If present, such a gap would be enhanced by magnetic field due to stronger confinement and increased exciton binding energy, contrary to our observations.

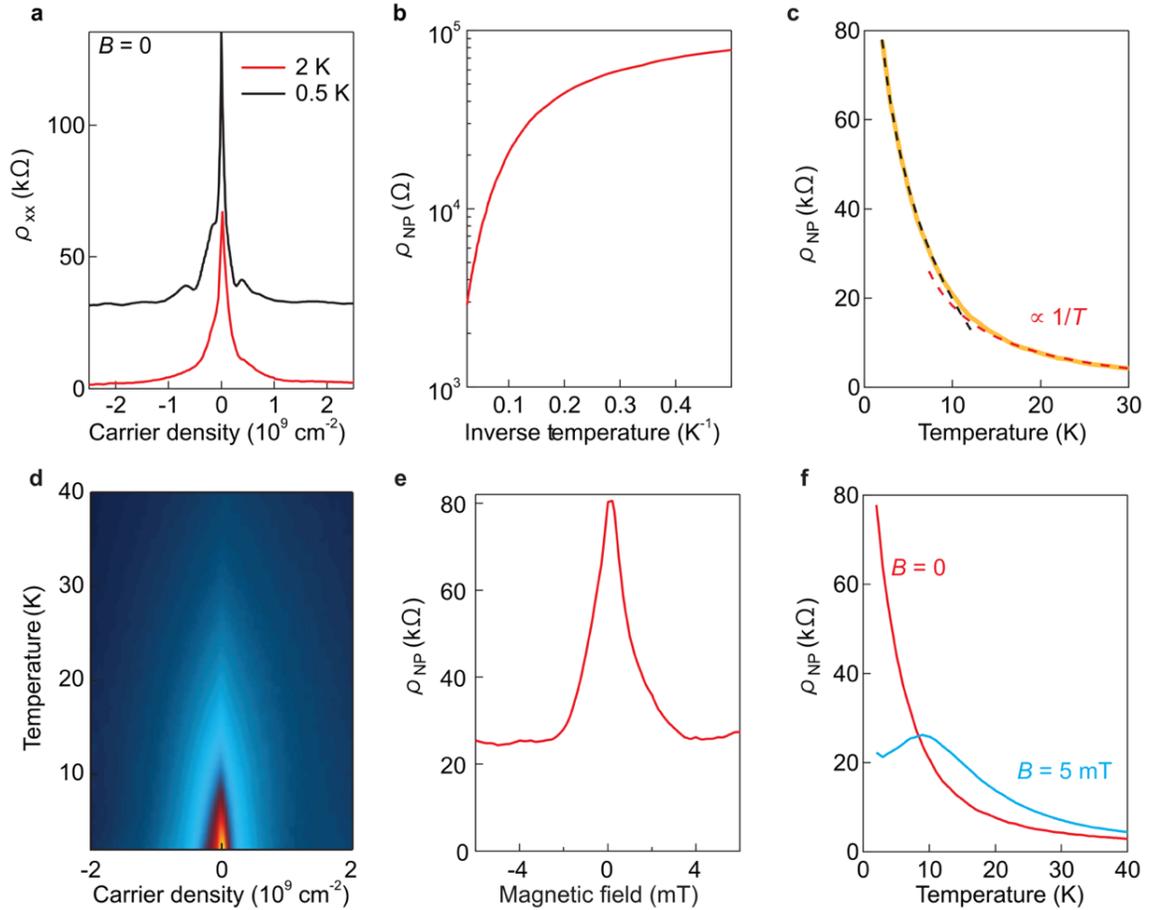

**Supplementary Fig. 4| Insulating behavior at the neutrality point. a** Zero-field $\rho_{xx}(n)$ at 0.5 and 2 K (color-coded curves offset by 30 kOhm for clarity). **b** Arrhenius plot of $\rho_{NP}(T)$. **c** Same data on linear scale (orange curve). Red dashed curve: Fit to $\rho_{NP} \propto 1/T$, characteristic of ballistic Dirac-fermion transport in the Boltzmann regime. Black curve: Logarithmic fit below 10 K. **d** Evolution of $\rho_{NP}(n,T)$ near the NP (dark blue to yellow: 0 to 80 kOhm). **e** $\rho_{NP}$ as a function of perpendicular $B$ at 2 K, showing suppression of the insulating state by small magnetic fields. **f** Temperature dependence of $\rho_{NP}$ at 0 and 5 mT (color-coded). The magnetoresistance changes sign around 10 K. Data for device S1 ($W \approx 9$ μm).

**#7 Magnetic focusing in reference devices**

Supplementary Figs. 5a,b present magneto-focusing measurements for hBN-encapsulated graphene devices with a Si-wafer gate (located at ~350 nm below graphene) and a graphite gate (~70 nm below), which are analogous to the measurements in Fig. 2 of the main text for proximity-gated devices. The focusing peaks appear at the expected positions given by[26,27] $B(n,P) = 2\hbar(\pi n)^{1/2}P/eL$ where $P$ is the peak number and $L$ the distance between injector and collector (dashed curves in Supplementary Figs. 5a,b).

We define the "first appearance" of magneto-focusing peaks as the carrier density $n$ where the $P = 1$ peak becomes clearly distinguishable above a noisy or fluctuating background, typically exceeding it by a factor of >2. Importantly, the peak grows very rapidly as a function of $n$, so that the exact threshold factor is relatively unimportant and, in our experience, led to variations of < 30% in the estimated onset density. Using this criterion, the magneto-focusing peaks in Supplementary Figs. 5a,b first appeared at $n \approx 0.5$ and $\approx 0.1 \times 10^{12}$ cm$^{-2}$ for the devices with Si and remote-graphite gates, respectively. For the measurement geometries shown in the figure insets with injector-collector separations $L \approx 19$ and 22.5 μm, this yields transport mobilities $\mu = (e\ell/\hbar)(\pi n)^{-1/2} \approx 2$ and $7 \times 10^6$ cm$^2$ V$^{-1}$ s$^{-1}$, respectively. They are in good agreement with the mobilities extracted from the devices' $\rho_{xx}(n)$ dependences at the same densities, which validates the suggested criterion for estimation of transport mobilities from the appearance of the first peak in magnetic focusing measurements. Accordingly, we use this criterion in the main text to estimate $\mu$ also in proximity-gated devices. The surprising accuracy of this rule-of-thumb criterion stems from the very rapid development of magneto-focusing signals with increasing $n$, which resembles the similarly simple criterion for estimations of quantum mobility as discussed in the next two sections. Supplementary Fig. 5c summarizes our findings for mobility in all three types of devices. Remote graphite gates improve $\mu$ several-fold compared to Si-gated devices (explaining why graphite gates have become common in the recent literature), whereas proximity screening provides an additional order-of-magnitude enhancement in electronic quality. Note that the achieved mobilities in Supplementary Fig. 5c have to be compared at different carrier densities because magnetic focusing becomes observable only at $n$ that exceeded $\delta n$ by 30–100 times, with $\delta n$ varying significantly between device types.

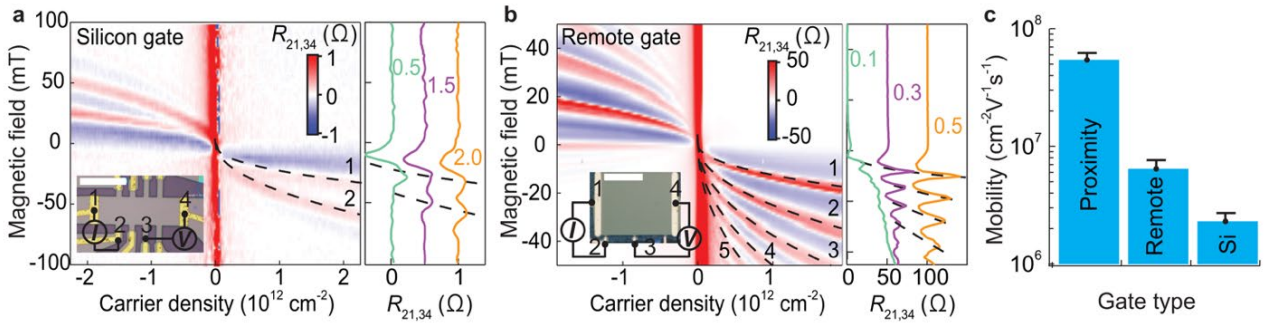

**Supplementary Fig. 5 | Magnetic focusing for different gate architectures. a** and **b** Measurements for reference devices with remote Si and graphite gates, respectively. Insets: Device micrographs and measurement configurations (current $I_{21}$ between contacts 2 and 1; voltage $V_{34}$ between 3 and 4; scale bars: 20 μm; $T = 20$ K). Color maps: Focusing resistance $R_{21,34}(n,B)$. Dashed curves indicate the expected position of magneto-focusing peaks (peak numbers $P$ are labelled). Right panels: Vertical cuts from the maps at the labelled densities (in units of $10^{12}$ cm$^{-2}$). **c** Transport mobilities found in our magneto-focusing experiments for devices with the three types of gates. The error bars indicate uncertainty in our evaluation of the extracted $\mu$.

#8 Onset of SdH oscillations

We studied seven graphene devices with proximity gates, and results for three of them with remnant $\delta n < 10^8$ cm$^{-2}$ are presented below. Supplementary Fig. 6 shows their Landau fan diagrams near the NP, along with horizontal cuts at several fields in the mT range. Quantization becomes clearly visible below 1, 2 and 3 mT for devices S2, S3 and S4, respectively, yielding $\mu_q$ exceeding 10, 5 and 3.3 ×10$^6$ cm$^2$V$^{-1}$s$^{-1}$, respectively. The quantum mobility in device S2 matches or exceeds that of device S1 in the main text. To ensure reliable identification of SdH oscillations, we considered them to emerge only if the resistivity minima at $\nu = \pm 2$ became pronounced enough to rule out any possible quantum-interference features (main text and section 'Metal-insulator transition'). This required fields notably higher than those needed for the criterion $\mu_q B^* \approx 1$, where SdH oscillations with amplitude less than $exp(-\pi) \approx 4\%$ of total $\rho_{xx}$ appear at $B^*$ (see the next section). Accordingly, the above values provide only lower bounds for $\mu_q$.

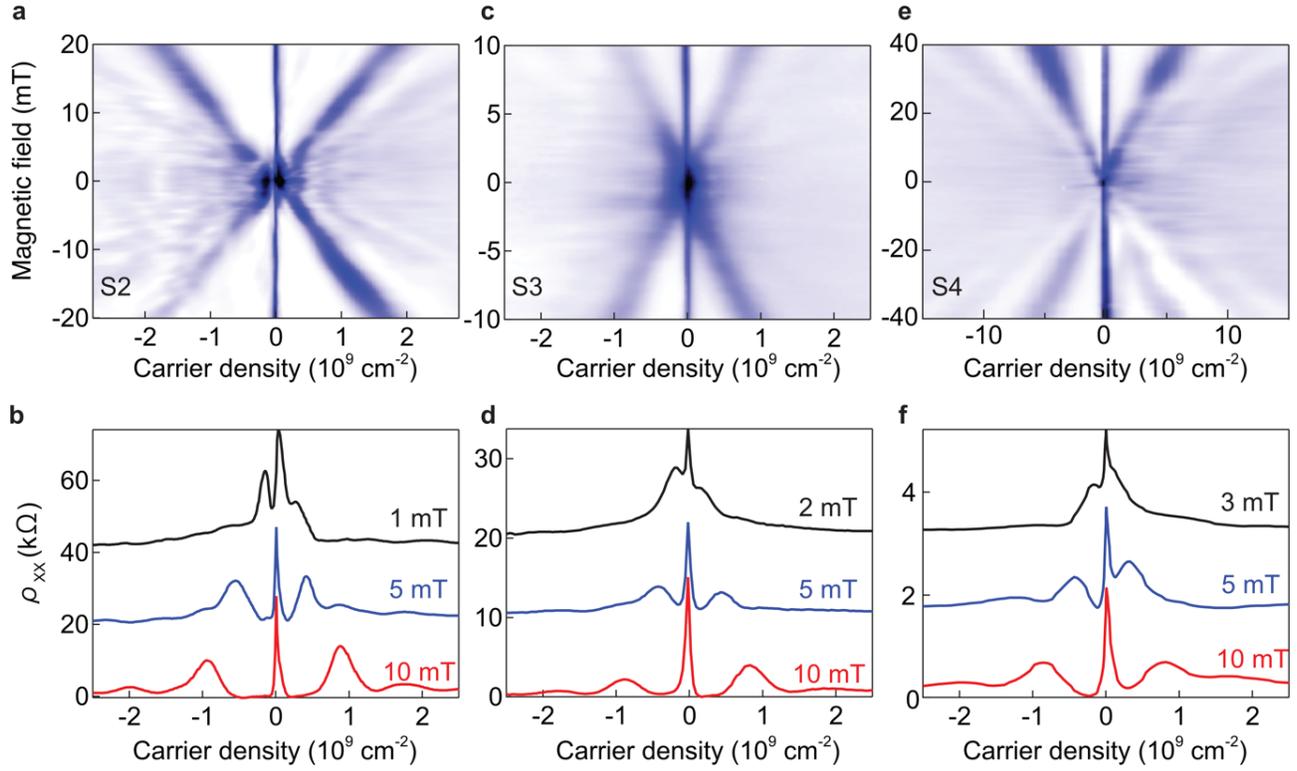

**Supplementary Fig. 6| Onset of SdH oscillations in different proximity-gated devices**. **a**, **c**, **e** Maps $\rho_{xx}(n,B)$ near the NP for devices S2, S3 and S4 (white-to-blue scales: 0 to 40, 25 and 3 kOhm, respectively). **b**, **d**, **f** Horizontal cuts from the maps above. All measurements at ~2 K. Note that the early SdH oscillations appear superimposed on the pronounced resistivity peak at the NP, creating a broad dark region below a few mT in the color maps **a**, **c**, **e** that obscures the onset of Landau quantization. The onset becomes apparent at lower fields using horizontal cuts as in **b**, **d**, **f**.

#9 Determination of quantum mobility

Although the expression $\mu_q B^* \approx 1$ is widely adopted in the literature, to justify its use in our study, we performed a quantitative analysis of SdH oscillations. To avoid complications arising from the metal-insulator transition, QHE and interference resonances, the measurements were carried out on devices with remote graphite gates at an elevated $T$ of 5 K. Under these conditions, SdH oscillations exhibit clear sinusoidal behavior, with their amplitude exponentially increasing with $B$ (examples are shown in Supplementary Fig. 7a).

To analyze such curves, we applied the Lifshitz–Kosevich (LK) formula[41]

$$\Delta\rho_{xx}(n,T,B) \propto \frac{2\pi^2 k_B T/\hbar\omega_c}{\sinh(2\pi^2 k_B T/\hbar\omega_c)} \exp(-\pi/\mu_q B)$$

where $\omega_c$ is the cyclotron frequency. For each carrier density, we replotted the oscillatory part of resistivity $\Delta\rho_{xx}$ as a function of $1/B$ and fitted it using the LK expression, with $\mu_q$ as the only fitting parameter (inset of Supplementary Fig. 7b). Supplementary Fig. 7b compares the quantum mobilities extracted from these fits with those obtained using the rule-of-thumb criterion $\mu_q B^* \approx 1$ with $B^*$ being the field at which SdH oscillations first become clearly discernible (empty diamonds in Supplementary Fig. 7a). Within our experimental accuracy (10–20%), the two methods yield identical values of $\mu_q$, validating our use of the simpler criterion.

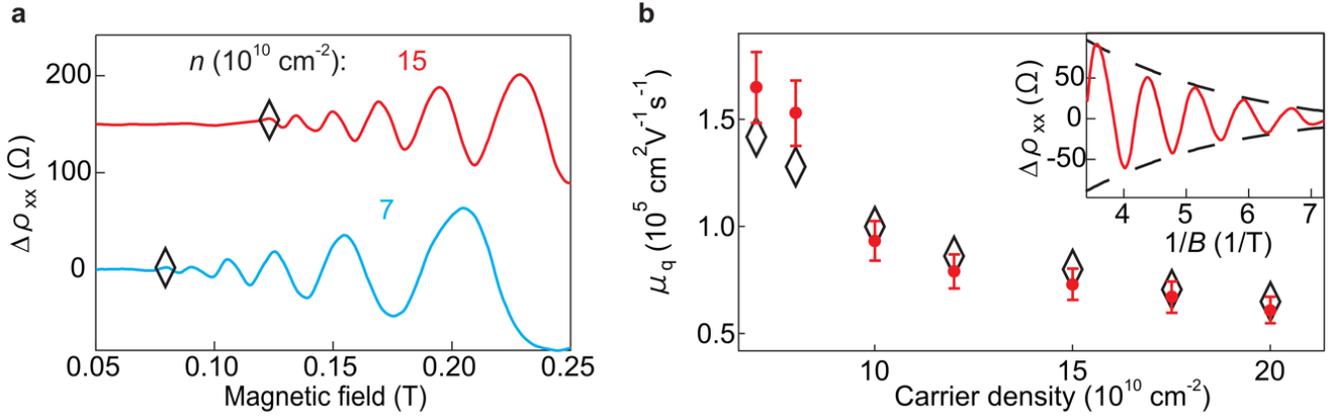

**Supplementary Fig. 7| Analysis of SdH oscillations' onset. a** Oscillatory part of resistivity $\Delta\rho_{xx}$ for two representative densities (smooth background subtracted; curves are offset for clarity). Empty diamonds mark the first visible oscillations. **b** Quantum mobilities extracted using the rule-of-thumb criterion for the onset of SdH oscillations (diamonds) and using LK theory fits (red symbols). Inset: Data from **a** replotted as a function of $1/B$. Dashed curves show LK fits through resistance maxima and minima.

#10 Onset of quantized Hall plateaus

Although in our proximity-gated devices quantum Hall plateaus appeared in fields as small as several mT, we found that contact geometry, rather than disorder, limited the QHE onset. Supplementary Fig. 8a shows Hall resistivity $\rho_{xy}(n)$ at different magnetic fields. As $B$ increases, peaks in $\rho_{xy}$ grow until reaching the quantized value of $h/2e^2$ at 4–5 mT. At these fields, cyclotron diameter $D_c$ becomes comparable to the width $w$ of our voltage probes (for Dirac fermions at $\nu = \pm 2$, $D_c = 2\ell_B \approx 1.6$ μm/$B$(mT)$^{1/2}$). Consequently, cyclotron orbits in lower $B$ become reflected, preventing edge states from reaching the contact regions and thus destroying quantization[29].

To corroborate this geometric effect, we examined the QHE onset in devices with different $w$. Supplementary Fig. 8b plots the onset field versus $w$ (blue symbols). Hall plateaus develop systematically at lower $B$ for wider contacts, following qualitatively the criterion $D_c \lesssim w$. This suggests that, in our proximity-gated devices, the onset of full quantization was still limited by voltage probe dimensions rather than graphene's homogeneity or mobility, and we expect the QHE in proximity-gated devices to fully develop in fields as low as 1 mT, if using contacts wider than ~2 μm.

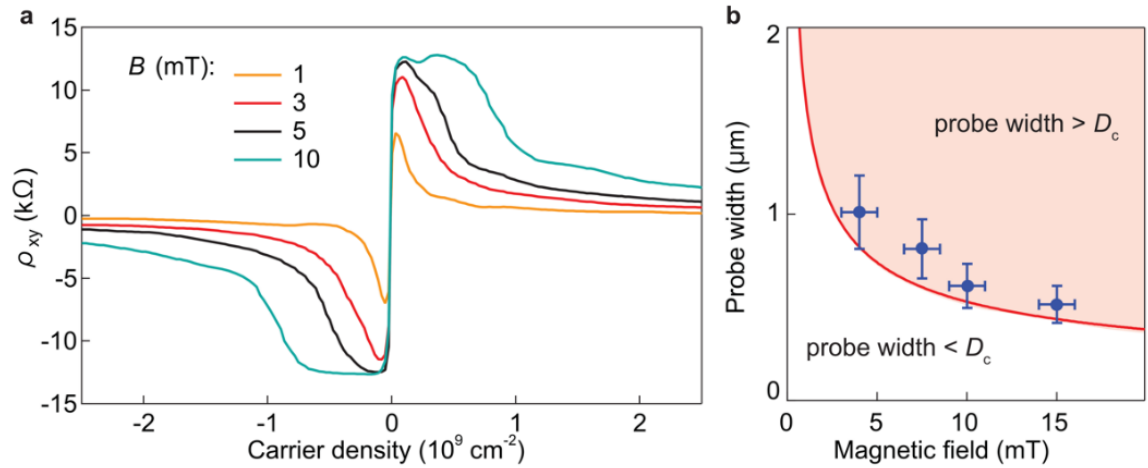

**Supplementary Fig. 8| Contact width and Hall quantization**. **a** Hall resistivity $\rho_{xy}(n)$ at different $B$ (color-coded curves) for device S1. Hall plateaus emerge below 5 mT. **b** Magnetic field required for their emergence versus voltage probes' width (blue symbols). Error bars indicate uncertainties in determining $w$ (from optical micrographs) and the quantization onset. Red curve: Criterion $D_c = w$. The pink area shows where $w$ exceeds $D_c$, allowing entry of edge states into ohmic contacts.

#11 Fractional quantum Hall effect under proximity screening

To assess how the fractional QHE develops in our proximity-gated devices we studied its detailed evolution with increasing $B$ as shown in Supplementary Fig. 9. No fractional states could be observed below 6 T. Their signatures at $\nu = 2/3$ and $5/3$ first appeared above 7 T and became more pronounced at higher fields (Supplementary Figs. 9a,b). Notably, no signatures of $\nu = 1/3$ were observed in any of our devices, which we attribute to the large negative quantum capacitance that increases the total capacitance near incompressible states and can lead to a complete collapse of visible gaps when measurements are performed as a function of gate voltage[31].

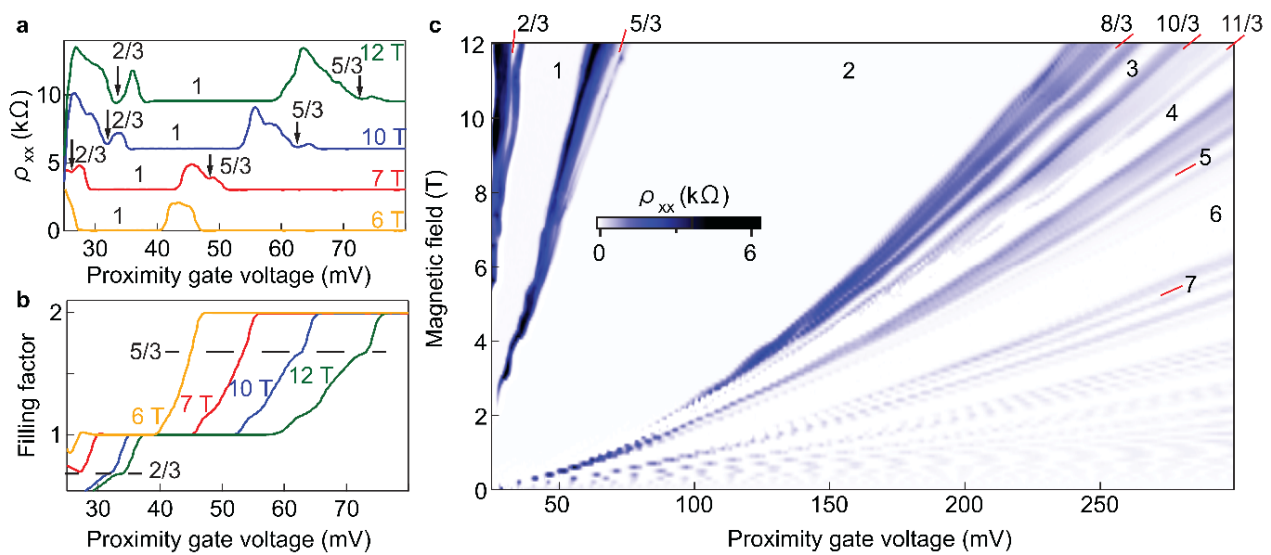

**Supplementary Fig. 9| Evolution of fractional quantum Hall states with magnetic field**. **a** and **b** $\rho_{xx}$ and $\nu = (h/e^2)/\rho_{xy}$ as functions of proximity gate voltage at different $B$ from 6 to 12 T. Vertical arrows indicate minima in $\rho_{xx}$. Horizontal dashed lines mark emerging Hall plateaus for $\nu = 2/3$ and $5/3$. **c** Map of $\rho_{xx}$ as a function of $B$ and proximity gate voltage. Numbers indicate different filling factors with red lines serving as guides to the eye.

The emergence of fractional states corresponds to the magnetic length $\ell_B$ decreasing below 10 nm. This is consistent with the fact that $\ell_B$ represents the spatial scale of electron-electron interactions responsible for the fractional QHE. This critical scale of 10 nm inferred from the emergence of fractional states at 7 T aligns well with the characteristic scale $2\pi\alpha \cdot d \approx 10$ nm at which proximity screening for $d \approx 1$ nm becomes effective in suppressing electron-hole puddles, as discussed in the main text. The systematic appearance of fractional QHE states only when $\ell_B < 10$ nm confirms that proximity screening primarily affects long-range interactions while preserving short-range ones essential for many-body phenomena in high magnetic fields.

**#12 Sources of disorder**

Despite considerable progress in improving the quality of graphene devices over the past decade, the primary sources of disorder limiting mobility and charge homogeneity in hBN-encapsulated graphene remain poorly understood. Our current experiments and the existing literature allow us to rule out several potential disorder sources. First, atomic-scale defects in graphene itself or at graphene-hBN interfaces are unlikely to be dominant scatterers. Scanning probe experiments would have revealed individual short-range defects that were clearly visible on the corresponding images[42] and did not occur within areas as large as >10 µm$^2$. The graphene-hBN interfaces are known to be atomically clean and flat, with minimal interfacial contamination[43] (except for bubbles and wrinkles, which however were absent in our devices; see Supplementary Information). Second, charged impurities at the metal-hBN interface of the top gate also appear to play a minimal role. Additional experiments demonstrated no significant improvement when using graphite for both top and bottom gates (placed at distances more than 10 nm). The latter observation is in line with previous studies using double-gated devices with two graphite gates, where mobilities were comparable to those in our remote-graphite-gated devices. Furthermore, attempts to use other atomically-flat metallic crystals as proximity gates (including $Bi_2Sr_2CaCu_2O_{8+x}$ and $TaS_2$) failed to yield high mobilities[21], presumably due to charged impurities at their surfaces in contact with hBN. Third, elastic strain in graphene can generate electron-hole puddles, but our observations suggest this to be an improbable primary source of disorder. Indeed, strain-induced puddles cannot be suppressed by proximity gating, yet we observed a dramatic reduction in charge inhomogeneity.

Beyond edge scattering, which obviously dominates in our limited-width devices, charged impurities within hBN represent the most likely source of both residual disorder and charge inhomogeneity. The hBN crystals used in our devices (from National Institute for Materials Science, Japan) contain impurities at concentrations of ~$10^{15}$ cm$^{-3}$, primarily carbon substituents (Takashi Taniguchi, private communication). In comparison, state-of-the-art GaAlAs heterostructures[8,9] contain impurities at concentrations of <$5\times10^{13}$ cm$^{-3}$. However, this comparison requires nuanced interpretation. Impurities in hBN typically lie deep within the bandgap and remain predominantly neutral, whereas impurities in GaAlAs heterostructures (such as Si used for doping) are mostly ionized. Nonetheless, it is plausible that a fraction of deep impurity states in hBN become charged, creating a background electrostatic potential in graphene. The observed effectiveness of proximity gating (particularly in suppressing charge inhomogeneity by two orders of magnitude) strongly supports this interpretation. Proximity screening specifically targets long-range electrostatic disorder while having minimal impact on short-range scattering mechanisms. While this analysis provides useful insights, a definitive conclusion about the dominant source(s) of disorder in hBN-encapsulated graphene demands further systematic investigation. This will probably require larger devices since the mobilities achieved in our proximity-gated devices have been limited by edge scattering due to their final size.

#13 Helical quantum Hall transport

While proximity screening suppresses electron-electron interactions, it can enable quantum phases that are otherwise difficult to observe in remote-gate devices. A notable example is the helical quantum Hall phase in the zeroth Landau level, which arises if a spin-polarized ferromagnetic phase is stabilized either by very large (~30 T) in-plane magnetic fields[44] or by screening Coulomb interactions using high-$\varepsilon$ substrates[36]. In standard graphene devices, the zeroth Landau level typically exhibits an antiferromagnetic ground state, resulting in insulating bulk and insulating edges. In contrast, the ferromagnetic phase leads to a quantum Hall topological insulator with an insulating bulk but conducting helical edge channels where current propagation direction is locked to electron spin polarization[36,44] (Supplementary Fig. 10a).

To test for helical edge transport in proximity-gated graphene, we measured two-terminal conductance in different contact configurations under perpendicular magnetic field. In our multiterminal Hall bar geometry, each ohmic contact acts as an equilibration reservoir for counter-propagating edge states and, therefore, each edge section between two contacts represents an ideal helical quantum conductor with resistance $h/e^2$. Considering conduction along both device edges, the two-terminal conductance can be written as $G_{2t} = (e^2/h)(N_R^{-1} + N_L^{-1})$ where $N_R$ and $N_L$ are the numbers of helical-conductor sections for the right and left edges, respectively (Supplementary Fig. 10a, right panel).

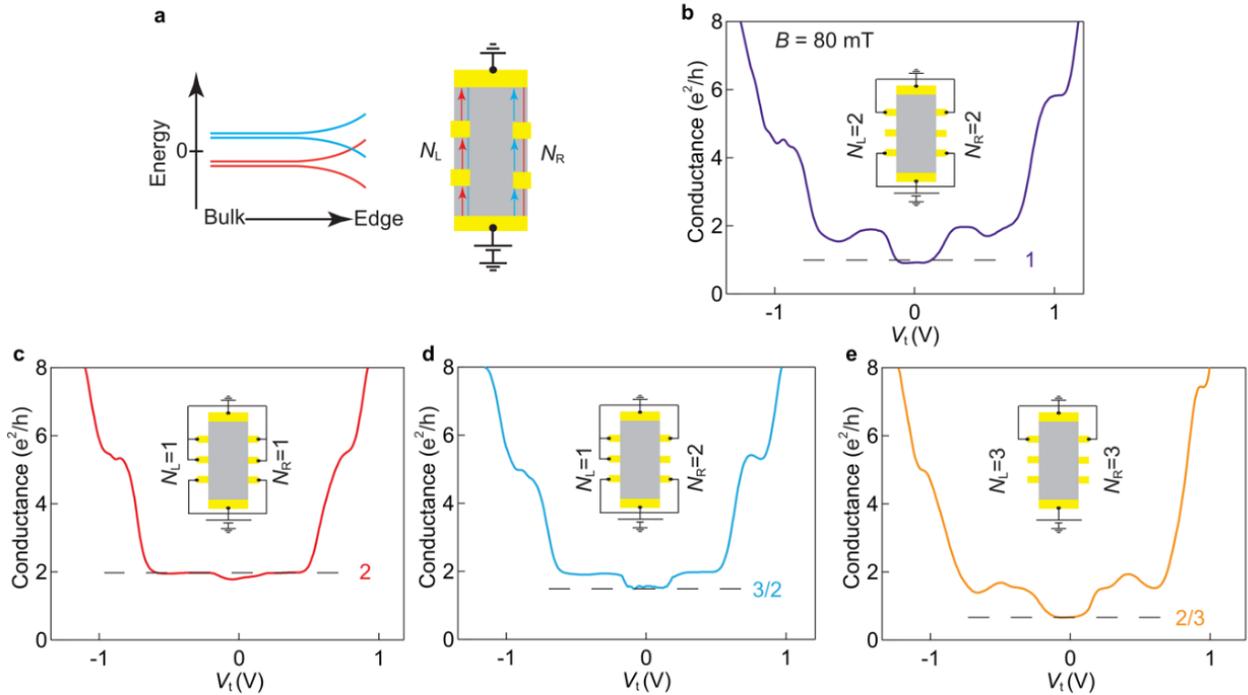

**Supplementary Fig. 10| Helical quantum Hall transport. a,** Left: Schematic of the zeroth Landau level in the ferromagnetic phase leading to helical edge states. Colors indicate spin-up and spin-down valley-degenerate Landau level pairs (red and blue, respectively). Right: Two-terminal measurement schematics. Yellow rectangles represent ohmic contacts. Arrows along edges indicate helical edge state propagation directions. $N_L$ and $N_R$ denote the numbers of helical-conductor sections at the left and right edges, respectively. **b-e** Two-terminal conductance $G_{2t}$ as a function of top gate voltage $V_t$ at 80 mT for different measurement configurations: $N_L = N_R = 2$ (**b**), $N_L = N_R = 1$ (**c**), $N_L = 1$ and $N_R = 2$ (**d**), and $N_L = N_R = 3$ (**e**). Dashed lines indicate the expected conductance values near $V_t = 0$ for each geometry. All measurements at 2 K.

Supplementary Fig. 10b shows the measured two-terminal conductance versus top gate voltage $V_t$ at $B$ = 80 mT, a relatively small but sufficient field to suppress the insulating state at the NP (see 'Metal-insulator transition').

As expected, well-defined conductance plateaus appear at $2e^2/h$ and $6e^2/h$, corresponding to the filling factors $\nu$ = ±2 and $\nu$ = ±6, respectively. Additionally, a quantized plateau emerges at the NP ($V_t$ = 0) with $G_{2t}$ = $e^2/h$, consistent with helical edge conduction and matching the expected value for our geometry with $N_L$ = $N_R$ = 2. Geometry-dependent measurements (Supplementary Figs. 10c-e) further support the presence of helical transport, where altering the number of contacts along the edges between the source and drain changed the conductance as expected from the above formula. In all configurations, the observed conductance at the NP showed quantized plateaus matching theoretical predictions.

These results demonstrate helical edge states in proximity-screened graphene at magnetic fields more than an order of magnitude lower than previously required (>1 T) using a high-$\varepsilon$ substrate[36]. This highlights the potential of proximity-gated devices for accessing exotic quantum Hall phases at reduced magnetic fields by suppressing electron interactions.

**Addendum**

**Graphene and hBN exfoliation:** The hBN and graphene crystals were obtained via mechanical exfoliation of bulk crystals using 'blue tape' ELP BT 130E-SL from *Nitto Denko*. The key steps in the exfoliation process are as follows:
1. Oxidized silicon substrates were cleaned by ultrasonic treatment in acetone for 5 minutes, followed by isopropyl alcohol (IPA) for another 5 minutes. After drying with nitrogen ($N_2$), the substrates were baked on a hotplate at 150 °C.
2. Graphite crystals (*NGS Naturgraphit*) were applied to the blue tape and cleaved 5–7 times, which optimized surface coverage while maintaining large lateral dimensions of individual crystals.
3. The tape with exfoliated thin graphite crystals was applied to the silicon substrate immediately after the substrate was removed from the hotplate. To further enhance exfoliation yield, we sometimes pretreated substrates with low-power $O_2$ plasma before exfoliation.
4. For hBN exfoliation, bulk crystals (National Institute for Materials Science, Tsukuba, Japan) were cleaved 5–8 times before applying the blue tape. To improve the exfoliation yield, the tape was pressed onto and then peeled from the substrate at 80 °C. Over the course of our experiments, we utilized several different batches of hBN crystals and found that the specific batch had no noticeable influence on the electronic quality of our proximity-gated devices.
5. To facilitate identification of ultrathin hBN flakes (~1 nm thick), we performed the exfoliation on Si wafers with a 70 nm $SiO_2$ layer.
6. Exfoliation and assembly were done in air. Exfoliated graphene and hBN crystals were either used as soon as possible (typically within one hour) or kept in a glove box for later usage within maximum 2-3 weeks.

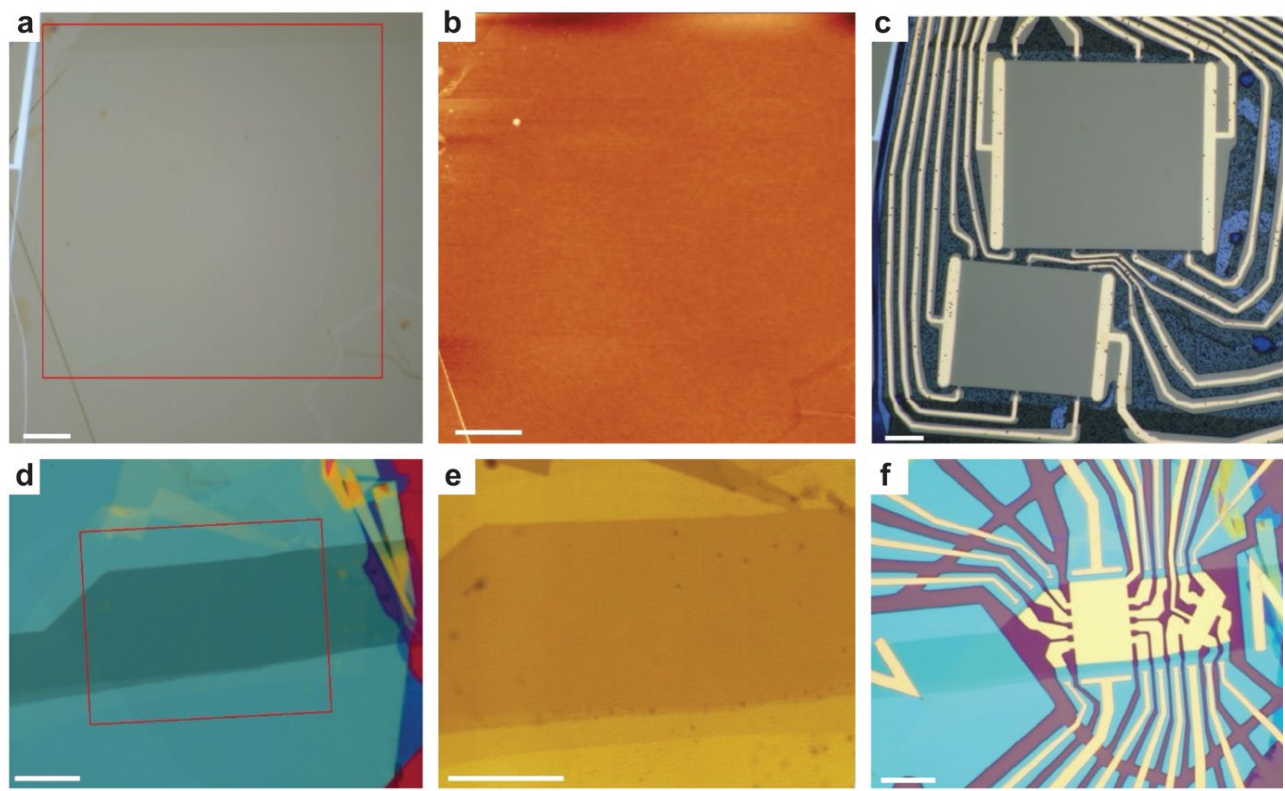

**Supplementary Fig. 1|** Optical micrographs of hBN-encapsulated graphene stacks used in devices with remote graphite gates (**a**) and with proximity gates (**d**), shown before microfabrication. Regions of interest are indicated by red boxes, with their AFM topography shown in panels **b** and **e**, whereas images of the final devices are shown in **c** and **f**, respectively. Scale bars, 10 μm for all panels.

**Assembly using PDMS stamps:** For proximity-gated devices and some of the reference devices, heterostructures were assembled using polydimethylsiloxane (PDMS) stamps coated with a layer of polypropylene carbonate (PPC). PDMS sheets (*Gel-Pak*, DGL-45X45-0170-X4, 1.7 mm thick) were cut into 5 mm × 5 mm squares, affixed to plasma-cleaned glass slides, and spin-coated with a PPC solution (1 g PPC in 10 mL chloroform) at 3,000 rpm. The prepared stamps were then annealed in air at 130 °C for 5 minutes before use in assembly procedures.

The assembly began with picking up the top hBN crystal, followed sequentially by graphene and the bottom hBN crystal, with the temperature held at 45 °C throughout all steps. The resulting trilayer stack was then deposited onto a graphite crystal (or a clean oxidized Si wafer for some reference devices) at 120 °C, which melted the PPC layer. PPC residues were removed using acetone, and the assembled heterostructure was subsequently annealed in vacuum at 250 °C for 2 hours for further cleaning.

To minimize the formation of hydrocarbon-filled bubbles and interfacial wrinkles during the assembly, we emphasize that 2D crystals should be picked up from substrates and deposited onto the target layer/substrate as slowly as possible. This slower transfer allowed contaminants to be driven outside the atomically-flat interface between the two assembled crystals. Supplementary Fig. 1 shows images of our devices with large areas free from bubbles and wrinkles, which enabled the fabrication of final extra-large devices.

**Assembly using silicon nitride membranes:** For some of our reference devices with remote graphite gates, we fabricated heterostructure stacks using polymer-free silicon nitride cantilevers. The assembly technique has been described previously[17] in extensive detail, including both the assembly process and cantilever fabrication. Here we provide further details and parameters specific to the heterostructure devices used in the present work.

Fresh silicon-nitride cantilevers were coated with a trilayer metal stack consisting of 1 nm of tantalum (Ta) for adhesion, 5 nm of platinum (Pt) to provide a smooth surface, and 0.75 nm of gold (Au) to promote adhesion with 2D materials. The top hBN crystal was picked up at 150 °C, followed sequentially by graphene and a bottom hBN layer (up to 70 nm thick) at 100 °C. The fully assembled hBN–graphene–hBN heterostructure was then released onto a large graphite flake at temperatures above 200 °C.

The entire assembly process was performed relatively rapidly, as elevated temperatures significantly enhanced the surface mobility of contaminant molecules, making the time of contact formation between 2D crystals less critical for avoiding contamination bubbles.

To ensure reliable release of the assembled stack from the cantilever, a substrate with high interfacial adhesion was necessary. Typically, thick (~50 nm) and laterally large (>100 μm) graphite crystals were used for this purpose, which also later served as remote gates. If graphite flakes were much thinner or smaller than heterostructure stacks, there was a high risk of failed release.

**Spectroscopic and microscopic characterization of heterostructures:** We employed Raman spectroscopy to verify the presence and determine the thicknesses of thin hBN crystals as well as to confirm monolayer graphene crystals. Atomic force microscopy (AFM) provided surface topography, enabling identification of regions free from interfacial bubbles, wrinkles, and cracks as shown in Supplementary Figs. 1**b**, **e**. Additionally, AFM was used to measure the thickness of individual layers within each stack.

**Microfabrication:** Following heterostructure assembly, electron-beam lithography was used to define the top gate region, followed by deposition of 2 nm Cr/50 nm Au using electron-beam evaporation. In a subsequent lithography step, electrical contacts were fabricated by first exposing graphene edges (using reactive-ion etching with a 90% $CHF_3$ – 10% $O_2$ gas mixture) and then depositing 2 nm Cr / 70 nm Au to form one-dimensional edge contacts. For the final step, the metallic top gate and additional lithographically defined links served as an etching mask to define multiterminal Hall bars, as shown in the main text and Supplementary Figs. 1c, d.